# Coherency vector formalism for polarimetric transformations

**Ignacio San José[1] and José J. Gil[2*]**

[1]*Instituto Aragonés de Estadística, Gobierno de Aragón, Bernardino Ramazzini 5, 50015 Zaragoza, Spain*
[2]*Departamento de Física Aplicada. Universidad de Zaragoza. Pedro Cerbuna 12, 50009 Zaragoza Spain*

*Corresponding author: ppgil@unizar.es

Despite the virtues of Jones and Mueller formalisms for the representation of the polarimetric properties, for some purposes in both Optics and SAR Polarimetry, the concept of coherency vector associated with a nondepolarizing medium has proven to be an useful mathematical structure that inherits certain symmetries underlying the nature of linear polarimetric transformations of the states of polarization of light caused by its interaction with material media. While the Jones and Mueller matrices of a serial combination of devices are given by the respective conventional matrix products, the composition of coherency vectors of such serial combinations requires a specific and unconventional mathematical rule. In this work, a vector product of coherency vectors is presented that satisfies, in a meaningful and consistent manner, the indicated requirements. As a result, a new algebraic formalism is built where the representation of polarization states of electromagnetic waves through Stokes vectors is preserved, while nondepolarizing media are represented by coherency vectors and general media are represented by coherency matrices generated by partially coherent compositions of the coherency vectors of the components.

## 1. Introduction

The transformations of states of polarization of light caused by linear interaction (refraction, reflection, scattering…) with material media can be represented mathematically by means of different structures like Jones matrices acting on Jones vectors or Mueller matrices acting on Stokes vectors [1,2]. The basic molecular interaction with incident electromagnetic waves has an intrinsic nondepolarizing nature, i.e. totally polarized states are transformed into totally polarized ones and can be represented through the Jones formalism. Nevertheless, partially polarized states cannot be described by means of Jones vectors and require more general structures like the ones given by Stokes vectors and polarization matrices. Further, in general, polarimetric measurements involve instrument-dependent statistical averages (temporal, spectral or spatial) [3] producing depolarization, in which case the Jones matrices are not longer applicable and the action of the medium has to be represented by Mueller matrices, coherency matrices, covariance matrices, or other structures, all of them involving up to sixteen independent parameters.

In this work, the peculiar features of the coherency vector to represent nondepolarizing media are analyzed and an original coherency product is defined in order to calculate directly the coherency vector of a serial composition of a number of nondepolarizing media represented by respective coherency vectors. This coherency vector algebraic approach has the virtue that the states of polarization remain represented by Stokes vectors. The general case of depolarizing media is then formulated by means of the partially coherent parallel composition of nondepolarizing coherency matrices (which in turn are defined in a simple and natural form in terms of the respective coherency vectors). From the biunivocal relation between coherency matrices and Mueller matrices, the said synthesis is also formulated in terms of Mueller matrices and illustrated by means of an example.

The relation of the coherency vector algebra with other polarization formalisms like the covariance vector and the matrix state is also described.

## 2. Theoretical background

There are many experimental circumstances where the interaction of totally polarized electromagnetic plane waves with a homogeneous and nondispersive linear medium, produces output beams that remain totally polarized. For such situations, any given totally polarized input state (hereafter called *pure state*) is characterized by a well-defined Jones vector $\boldsymbol{\varepsilon}$ [4,5], while the output pure state is characterized by a Jones vector $\boldsymbol{\varepsilon}'$. This kind of *nondepolarizing* linear interactions can be represented by the transformation $\boldsymbol{\varepsilon}' = \mathbf{T}\boldsymbol{\varepsilon}$, where the 2×2 complex matrix $\mathbf{T}$ is the so-called *Jones matrix* [4,5]. In that case, the medium as well as $\mathbf{T}$ and other mathematical structures representing it are said to be *nondepolarizing* (or *pure*).

Despite the fact that a given medium behaves as nondepolarizing, in general both the input and output electromagnetic beams can be partially polarized, in which case such states cannot be represented by Jones vectors, but can be rather represented by polarization matrices $\boldsymbol{\Phi}$, whose transformation by the medium, represented by $\mathbf{T}$, is given by [6]

$$\boldsymbol{\Phi}' = \left\langle \boldsymbol{\varepsilon}' \otimes \boldsymbol{\varepsilon}'^{\dagger} \right\rangle = \left\langle \mathbf{T}\boldsymbol{\varepsilon} \otimes \left(\mathbf{T}\boldsymbol{\varepsilon}\right)^{\dagger} \right\rangle = \mathbf{T} \left\langle \boldsymbol{\varepsilon} \otimes \boldsymbol{\varepsilon}^{\dagger} \right\rangle \mathbf{T}^{\dagger} = \mathbf{T}\boldsymbol{\Phi}\mathbf{T}^{\dagger}, \quad (1)$$

where $\left\langle \ \right\rangle$ denotes time, spectral or spatial ensemble averaging. An operational procedure for representing any kind of linear nondepolarizing transformation (1) is provided by the so-called *vectorial Pauli algebraic approach* [7-9].

Any Jones matrix can be submitted to the singular value decomposition $\mathbf{T} = \mathbf{T}_{R2}\,\mathrm{diag}(p_1, p_2)\,\mathbf{T}_{R1}$, where $\mathbf{T}_{R1}$ and $\mathbf{T}_{R2}$ are unitary matrices and $\mathbf{T}_{DL0} \equiv \mathrm{diag}(p_1, p_2)$ is a diagonal matrix whose nonzero elements are the real nonnegative singular values $p_1$ and $p_2$. As it is well known, the *principal amplitude coefficients* $p_1, p_2$ are the positive square roots of the nonnegative eigenvalues of matrix $\mathbf{T}\mathbf{T}^{\dagger}$, $\mathbf{T}_{R2}$ is the unitary matrix that diagonalizes $\mathbf{T}\mathbf{T}^{\dagger}$, and $\mathbf{T}_{R1}$ is the matrix that diagonalizes $\mathbf{T}^{\dagger}\mathbf{T}$.

The origin of the notation $\mathbf{T}_R$ used for unitary Jones matrices is because they represent birefringent media (retarders) [1].

The next subsections are devoted to the main peculiar features of Jones and Mueller formalisms, which will be reflected in the coherency formalism presented in further sections.

### 2.1 Passivity condition for Jones matrices

The condition for a Jones matrix to represent a passive pure medium arises from the physical restriction that the ratio between the intensities of the emerging and incident beams must be smaller than 1 for any input state of polarization. It is straightforward to show that this condition, called *passivity condition* (also *transmittance condition* or *gain condition*), can be formulated in terms of the largest principal intensity coefficient $p_1^2$ of $\mathbf{T}$ as $p_1^2 \leq 1$ [10,11]. By considering all passive Jones matrices proportional to $\mathbf{T}$, the one with minimal attenuation of intensity is given by the *canonical passive representative* Jones matrix $\hat{\mathbf{T}} \equiv \mathbf{T}/p_1$.

### 2.2 Product of Jones matrices

Given a pair of Jones matrices $\mathbf{T}_1$ and $\mathbf{T}_2$, the product $\mathbf{T}_2\mathbf{T}_1$ represents the successive serial actions of the media represented by them [4,5] so that the input Jones vector is transformed as $\mathbf{T}_2\mathbf{T}_1 \boldsymbol{\varepsilon} = \mathbf{T}_2 \boldsymbol{\varepsilon}' = \boldsymbol{\varepsilon}''$. The product of Jones matrices is not commutative, $\mathbf{T}_2\mathbf{T}_1 \neq \mathbf{T}_1\mathbf{T}_2$, which reflects the physical fact that the overall polarimetric effect of the serial actions of the media represented by $\mathbf{T}_1$ and $\mathbf{T}_2$ depends on their order in the serial arrangement.

### 2.3 Product of a Jones matrix and a scalar

Given a Jones matrix $\mathbf{T}$ and an arbitrary complex number $c \equiv |c|e^{i\varphi}$, the product $c\mathbf{T}$ can be expressed as $c\mathbf{T} = e^{i\varphi}\mathbf{T}_{R2}\,\mathrm{diag}(|c|p_1, |c|p_2)\,\mathbf{T}_{R1}$, so that, disregarding the overall phase factor $\exp(i\varphi)$, the multiplication is equivalent to the effect of an isotropic attenuator represented by $\mathbf{T}_c \equiv |c|\,\mathrm{diag}(1,1)$. Thus, provided $|c| \leq 1$, $\mathbf{T}_c$ and hence $c\mathbf{T}$, are passive Jones matrices, so that $c\mathbf{T}$ differs from $\mathbf{T}$ only by the overall attenuation and phase factors. If $|c| > 1$, then $\mathbf{T}_c$ is not passive-realizable and, regardless of whether $c\mathbf{T}$ is passive or not, $c\mathbf{T}$ cannot be interpreted as the serial combination $\mathbf{T}_c\mathbf{T}$ [1].

### 2.4 Additive composition of Jones matrices

Leaving aside passivity constraints, the sum of Jones matrices is a Jones matrix. From a physical point of view, the addition of a number $n$ of Jones matrices $\mathbf{T}_i$ $(i=1,...,n)$ should be performed through a linear combination $\sum_{i=1}^{n} c_i \mathbf{T}_i$ where the complex coefficients $c_i$ satisfy $\sum_{i=1}^{n} |c_i|^2 = 1$. The additive composition of Jones matrices applies if and only if all the electromagnetic waves exiting from the parallel components (represented by $\mathbf{T}_i$) are coherently composed, in which case any input pure polarization state is transformed into an output pure state.

### 2.5 Stokes-Mueller formalism

The link between the polarization matrix $\boldsymbol{\Phi}$ and the associated Stokes vector $\mathbf{s} \equiv (s_0, s_1, s_2, s_3)^T$ is given by the expansion [6,12]

$$\boldsymbol{\Phi} = \frac{1}{2}(s_0 \boldsymbol{\sigma}_0 + s_1 \boldsymbol{\sigma}_1 + s_2 \boldsymbol{\sigma}_2 + s_3 \boldsymbol{\sigma}_3) = \frac{1}{2}\begin{pmatrix} s_0 + s_1 & s_2 - is_3 \\ s_2 + is_3 & s_0 - s_1 \end{pmatrix}, \quad (2)$$

where $\boldsymbol{\sigma}_i$ is the set composed of the $2\times 2$ identity matrix $\boldsymbol{\sigma}_0$ together with the Pauli matrices (taken in the order that is commonly used in polarization optics, which differs from certain quantum formulations where the subscripts 2,3 are interchanged)

$$\boldsymbol{\sigma}_0 = \begin{pmatrix} 1 & 0 \\ 0 & 1 \end{pmatrix}, \ \boldsymbol{\sigma}_1 = \begin{pmatrix} 1 & 0 \\ 0 & -1 \end{pmatrix}, \ \boldsymbol{\sigma}_2 = \begin{pmatrix} 0 & 1 \\ 1 & 0 \end{pmatrix}, \ \boldsymbol{\sigma}_3 = \begin{pmatrix} 0 & -i \\ i & 0 \end{pmatrix}. \quad (3)$$

Stokes vectors (which obviously do not constitute a vector space) are characterized by the pair of inequalities $s_0 \geq 0$, $\mathbf{s}^T \mathbf{G} \mathbf{s} \geq 0$ where $\mathbf{G} \equiv \mathrm{diag}(1,-1,-1,-1)$ is the diagonal matrix associated with the Minkowski metric, $s_0$ is the intensity and the scalar nonnegative quantity $\Sigma \equiv \mathbf{s}^T \mathbf{G} \mathbf{s} = s_0^2 - s_1^2 - s_2^2 - s_3^2$ is called the *randomness* of the state represented by $\mathbf{s}$ [7], which is related to the degree of polarization $P$ by means of $P^2 = 1 - \Sigma/s_0^2$.

The elements of the polarization matrix $\boldsymbol{\Phi}$ ca be arranged as the components of the *polarization coherency vector*

$$\boldsymbol{\varphi} \equiv \langle \boldsymbol{\varepsilon} \otimes \boldsymbol{\varepsilon}^* \rangle = \left( \langle \varepsilon_1 \varepsilon_1^* \rangle, \langle \varepsilon_1 \varepsilon_2^* \rangle, \langle \varepsilon_2 \varepsilon_1^* \rangle, \langle \varepsilon_2 \varepsilon_2^* \rangle \right)^T, \quad (4)$$

which is related as follows with $\mathbf{s}$ [13]

$$\mathbf{s} = \sqrt{2}\mathcal{L}\boldsymbol{\varphi}, \quad \mathcal{L} \equiv \frac{1}{\sqrt{2}}\begin{pmatrix} 1 & 0 & 0 & 1 \\ 1 & 0 & 0 & -1 \\ 0 & 1 & 1 & 0 \\ 0 & i & -i & 0 \end{pmatrix}, \quad (5)$$

where the matrix $\mathcal{L}$ connecting $\mathbf{s}$ and $\boldsymbol{\varphi}$ is unitary, $\mathcal{L}^{-1} = \mathcal{L}^\dagger$.

The transformation (1) of the state $\boldsymbol{\Phi}$ by the interaction with a nondepolarizing medium represented by the Jones matrix $\mathbf{T}$, can alternatively expressed in the form

$$\begin{aligned}\boldsymbol{\varphi}' &= \langle \boldsymbol{\varepsilon}' \otimes \boldsymbol{\varepsilon}'^* \rangle = \langle (\mathbf{T}\boldsymbol{\varepsilon}) \otimes (\mathbf{T}\boldsymbol{\varepsilon})^* \rangle \\ &= \langle (\mathbf{T} \otimes \mathbf{T}^*)(\boldsymbol{\varepsilon} \otimes \boldsymbol{\varepsilon}^*) \rangle = (\mathbf{T} \otimes \mathbf{T}^*)\boldsymbol{\varphi},\end{aligned} \quad (6)$$

and by combining Eqs. (5) and (6) we get

$$\mathbf{s}' = \sqrt{2}\mathcal{L}\boldsymbol{\varphi}' = \sqrt{2}\mathcal{L}(\mathbf{T} \otimes \mathbf{T}^*)\boldsymbol{\varphi} = \mathcal{L}(\mathbf{T} \otimes \mathbf{T}^*)\mathcal{L}^\dagger \mathbf{s}, \quad (7)$$

where the *pure Mueller matrix* (also *nondepolarizing* or *Mueller-Jones matrix*) transforming $\mathbf{s}$ into $\mathbf{s}'$ is defined from $\mathbf{T}$ as

$$\mathbf{M}(\mathbf{T}) \equiv \mathcal{L}(\mathbf{T} \otimes \mathbf{T}^*)\mathcal{L}^\dagger. \quad (8)$$

Following a common convention in polarization optics, wherever appropriate we will use the subscript $J$ to indicate that $\mathbf{M}_J$ is a pure Mueller matrix, i.e., $\mathbf{M}_J$ is derivable from a single Jones matrix as in Eq. (8). Subscript $J$ will be likewise used in other kinds of matrices when they represent nondepolarizing systems. Note that the conversion from $\mathbf{T}$ to $\mathbf{M}_J$ in Eq. (8) produces the removal of any global phase factor $e^{i\phi}$ affecting $\mathbf{T}$. Except for the cases where $\mathbf{T}$ is coherently composed with other Jones matrices, $\phi$ is non-measurable and therefore the factor $e^{i\phi}$ can be neglected, so that a pure Mueller matrix provides complete information on how a given nondepolarizing linear medium transforms an arbitrary polarization state.

The elements of $\mathbf{M}_J$, denoted as $m_{ij}$, are real and can be expressed in terms of its associated $\mathbf{T}$ as

$$m_{ij} = \frac{1}{2}\mathrm{tr}\left(\boldsymbol{\sigma}_i \mathbf{T} \boldsymbol{\sigma}_j \mathbf{T}^\dagger\right). \quad (9)$$

A real 4×4 matrix is a pure Mueller matrix if, and only if, it can be expressed as in Eq. (8) in terms of a Jones matrix. Moreover, any pure Mueller matrix $\mathbf{M}_J$ satisfies the characteristic property [14]

$$\mathbf{G}\mathbf{M}_J^T \mathbf{G}\mathbf{M}_J = \sqrt{\det \mathbf{M}_J}\, \mathbf{I}_4, \quad \begin{bmatrix} \mathbf{G} \equiv \mathrm{diag}(1,-1,-1,-1) \\ \mathbf{I}_4 \equiv \mathrm{diag}(1,1,1,1) \\ \sqrt{\det \mathbf{M}_J} = (p_1 p_2)^4 \end{bmatrix}. \quad (10)$$

The above result shows that the set of pure Mueller matrices is not completely characterized by the fact that they transform pure Stokes vectors into pure Stokes vectors. Additional constraints are embedded into the definition (8) of pure Mueller matrix [15].

Leaving aside a global phase factor of **T** (which is unphysical and negligible except for the case that **T** is additively composed with other Jones matrices), there is a biunivocal relation between a given Jones matrix **T** and its corresponding pure Mueller matrix $\mathbf{M}_J$, and both structures depend on up to seven independent measurable quantities. The redundancy of parameters in $\mathbf{M}_J$ (sixteen parameters to represent seven independent ones) is justified because the general case of depolarizing systems involves up to sixteen independent parameters of the composed (nonpure) Mueller matrix and therefore the Jones formalism is no longer applicable.

The polarimetric transformation of the Stokes vector **s** of an input polarization state into the Stokes vector **s**′ of an output polarization state by the effect of a linear passive medium can be represented, in general, as $\mathbf{s}' = \mathbf{M}\mathbf{s}$, where the $4\times 4$ real matrix **M** is the so-called Mueller matrix, which in general is depolarizing. As will be shown in further sections, the characterization and properties of general depolarizing linear media can be formulated from the appropriate additive compositions of nondepolarizing media.

Any Mueller matrix (pure or nonpure) can be expressed in a convenient partitioned form that simplifies the treatment of certain interesting properties [14],

$$\mathbf{M} = m_{00}\hat{\mathbf{M}}, \quad \hat{\mathbf{M}} \equiv \begin{pmatrix} 1 & \mathbf{D}^T \\ \mathbf{P} & \mathbf{m} \end{pmatrix}, \quad (11)$$

where the three-component vectors **D** and **P** are respectively called the *diattenuation vector* and the *polarizance vector* of **M** [16]. The magnitudes of these vectors are the *diattenuation*, $D \equiv |\mathbf{D}|$ and the *polarizance*, $P \equiv |\mathbf{P}|$ [17,18]. From the general expression (8) of a pure Mueller matrix as a function of its corresponding Jones matrix, it is straightforward to find that pure Mueller matrices satisfy the equality $P = D$ [11]. Nevertheless, this equality is not a necessary condition for general (depolarizing) Mueller matrices. Further, the mean intensity coefficient of **M** (i.e., intensity transmittance or reflectance for unpolarized input states) is given by $m_{00}$. Another interesting quantity that, as with $D$ and $m_{00}$ is invariant under arbitrary rotations of the coordinate system [19] is the degree of spherical purity, defined as $P_s(\mathbf{M}) \equiv \sqrt{\|\mathbf{m}\|_2^2 / 3}$ ($\|\mathbf{m}\|_2$ being the Frobenius norm of the submatrix **m** of **M**) [20].

Given a set of pure Mueller matrices $\mathbf{M}_{Ji}$ $(i=1,...,n)$, the product $\mathbf{M}_{Jn}...\mathbf{M}_{Ji}...\mathbf{M}_{J1}$ is a pure Mueller matrix that represents the successive serial actions of the media associated with them.

Given a pure Mueller matrix $\mathbf{M}_J$ and an arbitrary real positive number $c$, the product $c\mathbf{M}_J$ can be expressed as the matrix product [1]

$$c\mathbf{M}_J = (c\mathbf{I}_4)\mathbf{M}_J = \mathbf{M}_{R2}\mathbf{M}_{DL0}(c p_1^2, c p_2^2)\mathbf{M}_{R1}, \quad (12)$$

where $\mathbf{I}_4$ is the $4\times 4$ identity matrix and $\mathbf{M}_{DL0}(c p_1^2, c p_2^2)$ is the Mueller matrix of a normal linear diattenuator oriented at 0° with principal intensity coefficients $c p_1^2$ and $c p_2^2$ [1], so that the multiplication is equivalent to the effect of the neutral attenuating Mueller matrix $\mathbf{M}_{Jc} \equiv c\mathbf{I}_4$. Thus, provided $c \leq 1$, $\mathbf{M}_{Jc}$ and hence $c\mathbf{M}_J$, are passive Mueller matrices, and $c\mathbf{M}_J$ only differs from $\mathbf{M}_J$ by the overall attenuation factor $c$. For values $c > 1$, $\mathbf{M}_{Jc}$ is not passive.

*2.6 Additive composition of Mueller matrices*

Leaving aside passivity constraints, the sum of Mueller matrices is a Mueller matrix (which is depolarizing except for the trivial case where the addends are mutually proportional). From a physical point of view, the addition of a number $n$ of Mueller matrices $\mathbf{M}_i$ $(i=1,...,n)$ should be performed through a convex linear combination $\sum_{i=1}^n p_i \mathbf{M}_i$, with $p_i \geq 0$ and $\sum_{i=1}^n p_i = 1$ [15,21].

## 3. The coherency vector of a nondepolarizing medium

By expanding the Jones matrix **T** in terms of the Pauli matrices,

$$\mathbf{T} \equiv \begin{pmatrix} t_0 & t_1 \\ t_2 & t_3 \end{pmatrix} = c_0\boldsymbol{\sigma}_0 + c_1\boldsymbol{\sigma}_1 + c_2\boldsymbol{\sigma}_2 + c_3\boldsymbol{\sigma}_3$$

$$= \begin{pmatrix} c_0 + c_1 & c_2 - ic_3 \\ c_2 + ic_3 & c_0 - c_1 \end{pmatrix}, \quad (13)$$

$$c_0 = (t_0 + t_3)/2, \; c_1 = (t_0 - t_3)/2,$$

$$c_2 = (t_1 + t_2)/2, \; c_3 = i(t_1 - t_2)/2,$$

the complex coefficients $c_i$ can be arranged as the components of the *coherency vector* $\mathbf{c} = (c_0, c_1, c_2, c_3)^T$, while the corresponding *pure coherency matrix* $\mathbf{C}_J$ associated with **T** and $\mathbf{M}_J$, is defined as $\mathbf{C}_J = \mathbf{c}\otimes \mathbf{c}^\dagger$. Form its very definition, $\mathbf{C}_J$ is Hermitian, positive semidefinite, with $\text{rank}\,\mathbf{C}_J = 1$, and has peculiar and interesting features, as for instance the fact that it is diagonal if and only if $\mathbf{M}_J$ is diagonal, so that the use of **c** and $\mathbf{C}_J$ results advantageous when dealing with certain properties of Mueller matrices.

The matrix state **Z** (or simply the **Z**-matrix) is defined from **c** as follows [21,23]

$$\mathbf{Z} = \begin{pmatrix} c_0 & c_1 & c_2 & c_3 \\ c_1 & c_0 & -ic_3 & ic_2 \\ c_2 & ic_3 & c_0 & -ic_1 \\ c_3 & -ic_2 & ic_1 & c_0 \end{pmatrix} = \mathcal{L}(\mathbf{T}\otimes\mathbf{I}_2)\mathcal{L}^\dagger, \quad (14)$$

so that the corresponding Jones and Mueller matrices can be expressed as

$$\mathbf{T} = \begin{pmatrix} z_{00} + z_{01} & z_{02} - iz_{03} \\ z_{02} + iz_{03} & z_{00} - z_{01} \end{pmatrix},$$

$$\mathbf{M}_J = \mathbf{Z}\mathbf{Z}^* = \mathbf{Z}^*\mathbf{Z}. \quad (15)$$

As with **T**, **Z** depends on up to eight independent parameters, global phase factor included [22,23].

## 4. Product composition of coherency vectors

Obviously, there is no conventional mathematical operation associated with the serial composition of pure elements where the ordered "product" of their coherency vectors provides the coherency vector of the composed system. Nevertheless, such composition rule can be achieved by defining the following operation (hereafter called *coherency product*, and represented by the symbol •). Let **u** and **v** be two coherency vectors, the coherency product **u**•**v** is defined as

$$\mathbf{u}\bullet\mathbf{v} = \begin{pmatrix} \mathbf{u}^T \mathbf{G}_0 \mathbf{v} \\ \mathbf{u}^T \mathbf{G}_1 \mathbf{v} \\ \mathbf{u}^T \mathbf{G}_2 \mathbf{v} \\ \mathbf{u}^T \mathbf{G}_3 \mathbf{v} \end{pmatrix}, \quad \mathbf{G}_0 \equiv \mathbf{I}_4,$$

$$\mathbf{G}_1 \equiv \begin{pmatrix} 0 & 1 & 0 & 0 \\ 1 & 0 & 0 & 0 \\ 0 & 0 & 0 & i \\ 0 & 0 & -i & 0 \end{pmatrix}, \mathbf{G}_2 \equiv \begin{pmatrix} 0 & 0 & 1 & 0 \\ 0 & 0 & 0 & -i \\ 1 & 0 & 0 & 0 \\ 0 & i & 0 & 0 \end{pmatrix}, \mathbf{G}_3 \equiv \begin{pmatrix} 0 & 0 & 0 & 1 \\ 0 & 0 & i & 0 \\ 0 & -i & 0 & 0 \\ 1 & 0 & 0 & 0 \end{pmatrix}, \quad (16)$$

where $\mathbf{G}_0$ is the 4×4 identity matrix $\mathbf{I}_4$, $\mathbf{G}_1$, $\mathbf{G}_2$ and $\mathbf{G}_3$ are traceless, while the four $\mathbf{G}_i$ satisfy the following properties, which are analogous to those of the Pauli matrices [24]

$$\det \mathbf{G}_i = 1, \ \mathbf{G}_i^\dagger = \mathbf{G}_i^{-1}, \ \mathbf{G}_i^\dagger = \mathbf{G}_i,$$

$$\begin{aligned}
\mathbf{G}_1 \mathbf{G}_2 &= -i\mathbf{G}_3, \ \mathbf{G}_2 \mathbf{G}_3 = -i\mathbf{G}_1, \ \mathbf{G}_3 \mathbf{G}_1 = -i\mathbf{G}_2, \\
\mathbf{G}_2 \mathbf{G}_1 &= i\mathbf{G}_3, \ \mathbf{G}_3 \mathbf{G}_2 = i\mathbf{G}_1, \ \mathbf{G}_1 \mathbf{G}_3 = i\mathbf{G}_2.
\end{aligned} \quad (17)$$

Note that the components of the coherency vector $\mathbf{u} \bullet \mathbf{v}$ are given by $(\mathbf{u} \bullet \mathbf{v})_i = \mathbf{u}^T \mathbf{G}_i \mathbf{v}$.

It is straightforward to prove that, in analogy to what happens with Jones and Mueller matrices, given two nondepolarizing media with respective coherency vectors $\mathbf{u}$ and $\mathbf{v}$, the coherency vector of the serial combination of such media is precisely $\mathbf{w} = \mathbf{u} \bullet \mathbf{v}$. In other words, given two Jones matrices, $\mathbf{T}_u$ and $\mathbf{T}_v$ with associated Mueller matrices $\mathbf{M}_u$ and $\mathbf{M}_v$ and coherency vectors $\mathbf{u}$ and $\mathbf{v}$, respectively, the medium to which corresponds the product $\mathbf{T}_w = \mathbf{T}_u \mathbf{T}_v$ has an associated Mueller matrix $\mathbf{M}_{Jw} = \mathbf{M}_{Ju} \mathbf{M}_{Jv}$ and an associated coherency vector $\mathbf{w} = \mathbf{u} \bullet \mathbf{v}$. Therefore, the coherency product defined above provides the desired mathematical composition rule and replicates the properties of the product of Jones matrices, i.e. non-commutative, associative, etc. Furthermore, from the very definition of the coherency vector associated with a Jones matrix, the properties of the product by a scalar are also directly inherited.

The elements $m_{ij}$ of the Mueller matrix $\mathbf{M}_J$ associated with a given coherency vector $\mathbf{c}$ can be expressed in the simple form

$$m_{ij} = \mathbf{c}^\dagger \left(\mathbf{G}_i^* \mathbf{G}_j\right) \mathbf{c}, \quad (18)$$

showing that each element $m_{ij}$ is given by a norm (or measure) of the coherency vector $\mathbf{c}$, through the corresponding metric $\mathbf{G}_i^* \mathbf{G}_j$.

The transformation $\mathbf{s}' = \mathbf{M}_J \mathbf{s}$ of an input pure or nonpure state, into an output one with respective Stokes vectors $\mathbf{s}$ and $\mathbf{s}'$, is expressed in terms of the coherency vector $\mathbf{c}$ associated with $\mathbf{M}_J$ in the compact manner $\mathbf{s}' = \mathbf{c} \bullet \mathbf{s} \bullet \mathbf{c}^*$. Another interesting property is that the randomness $\Sigma$ of $\mathbf{s}$ is transformed as $\Sigma' = |\mathbf{c}^T \mathbf{G} \mathbf{c}|^2 \Sigma = \sqrt{\det \mathbf{M}_J} \, \Sigma$.

## 5. Properties of the coherency product

Below, we summarize the main properties of the coherency product defined in Eq. (16)

$\forall \mathbf{u}, \mathbf{v} \in \mathbb{C}^4, \mathbf{u} \bullet \mathbf{v} \in \mathbb{C}^4$,

Given $\mathbf{u}, \mathbf{v}, \mathbf{w} \in \mathbb{C}^4$, then $(\mathbf{u} \bullet \mathbf{v}) \bullet \mathbf{w} = \mathbf{u} \bullet (\mathbf{v} \bullet \mathbf{w})$,

$\forall \mathbf{c} \in \mathbb{C}^4 \ \exists \ \mathbf{e}_0 \ni \mathbf{c} \bullet \mathbf{e}_0 = \mathbf{e}_0 \bullet \mathbf{c}$ with $\mathbf{e}_0^T = (1,0,0,0)$,

$\forall \mathbf{c} \in \mathbb{C}^4 \ni \mathbf{c}^T \mathbf{G} \mathbf{c} \neq 0, \exists \ \mathbf{c}^{-1} \in \mathbb{C}^4 \ni \mathbf{c} \bullet \mathbf{c}^{-1} = \mathbf{c}^{-1} \bullet \mathbf{c} = \mathbf{e}_0$,

$\mathbf{c}^{-1} = \mathbf{G}\mathbf{c} / (\mathbf{c}^T \mathbf{G} \mathbf{c})$,

$\mathbf{G}(\mathbf{u} \bullet \mathbf{v}) = \mathbf{G}\mathbf{v} \bullet \mathbf{G}\mathbf{u}$,

$(\mathbf{u} \bullet \mathbf{v})^{-1} = \mathbf{v}^{-1} \bullet \mathbf{u}^{-1}, \ (\mathbf{u} \bullet \mathbf{v})^* = \mathbf{v}^* \bullet \mathbf{u}^*$,

Given $\mathbf{u}, \mathbf{v} \in \mathbb{C}^4$ and $p, q \in \mathbb{C}$, then $p\mathbf{u} \bullet q\mathbf{v} = pq(\mathbf{u} \bullet \mathbf{v})$,

Given $\mathbf{a}, \mathbf{b}, \mathbf{c}, \mathbf{d} \in \mathbb{C}^4$, then $(\mathbf{a} + \mathbf{b}) \bullet (\mathbf{c} + \mathbf{d}) = \mathbf{a} \bullet \mathbf{c} + \mathbf{a} \bullet \mathbf{d} + \mathbf{b} \bullet \mathbf{c} + \mathbf{b} \bullet \mathbf{d}$,

The coherency product is non-commutative, $\mathbf{u} \bullet \mathbf{v} \neq \mathbf{v} \bullet \mathbf{u}$,

$(\mathbf{u} \bullet \mathbf{v})^T \mathbf{G}(\mathbf{u} \bullet \mathbf{v}) = (\mathbf{u}^T \mathbf{G} \mathbf{u})(\mathbf{v}^T \mathbf{G} \mathbf{v})$,

$(\mathbf{w} \bullet \mathbf{u})^T \mathbf{G}(\mathbf{w} \bullet \mathbf{v}) = (\mathbf{w}^T \mathbf{G} \mathbf{w})(\mathbf{u}^T \mathbf{G} \mathbf{v})$,

$(\mathbf{u} \bullet \mathbf{v})^T \mathbf{w} = (\mathbf{w} \bullet \mathbf{u})^T \mathbf{v} = (\mathbf{v} \bullet \mathbf{w})^T \mathbf{u}$,

$\mathbf{c} \bullet \mathbf{e}_i = \mathbf{G}_i \mathbf{c}$ with $\begin{bmatrix} \mathbf{e}_0^T = (1,0,0,0), \ \mathbf{e}_1^T = (0,1,0,0) \\ \mathbf{e}_2^T = (0,0,1,0), \ \mathbf{e}_3^T = (0,0,0,1) \end{bmatrix}$,

$\mathbf{e}_i \bullet \mathbf{c} = \mathbf{G}_i^T \mathbf{c} = \mathbf{G}_i^* \mathbf{c}$,

$(\mathbf{c}^* \bullet \mathbf{e}_i)^T (\mathbf{c} \bullet \mathbf{e}_j) = \mathbf{c}^\dagger \mathbf{G}_i^* \mathbf{G}_j \mathbf{c} \ \ (i = 0,1,2,3)$,

$\forall \mathbf{c} \in \mathbb{C}^4 \ \mathbf{c}^* \bullet \mathbf{c}$ and $\mathbf{c} \bullet \mathbf{c}^*$ are Stokes vectors,

$\det \mathbf{M}_J(\mathbf{c}) = |\det \mathbf{T}(\mathbf{c})|^4 = |\mathbf{c}^T \mathbf{G} \mathbf{c}|^4$,

$\forall \mathbf{c} \in \mathbb{C}^4 \ni \mathbf{c}^T \mathbf{G} \mathbf{c} \neq 0$, vectors $\mathbf{G}_i \mathbf{u}$ are linearly independent,

$\mathbf{M}(\mathbf{G}\mathbf{u} \bullet \mathbf{v}) = \mathbf{G} \, \mathbf{M}(\mathbf{u}^T) \mathbf{G} \mathbf{M}(\mathbf{v}) = \mathbf{G} \, \mathbf{M}(\mathbf{u}^*) \mathbf{G} \mathbf{M}(\mathbf{v})$,

$\left[\text{with } \mathbf{M}(\mathbf{G}\mathbf{u}) = \mathbf{G} \mathbf{M}(\mathbf{u}^*) \mathbf{G}\right]$.

## 6. Formulation of the polarimetric properties of nondepolarizing media through the coherency vector approach

Concerning the properties of nondepolarizing media, usually expressed in terms of Mueller or Jones matrices, they have simple and interesting formulations in terms of the coherency vector $\mathbf{c}$. Some of these properties are summarized below.

The mean intensity coefficient is given by the square of the absolute value of $\mathbf{c}$,

$$|\mathbf{c}|^2 = \mathbf{c}^\dagger \mathbf{c} = m_{00} = \|\mathbf{T}\|_2^2. \quad (19)$$

The principal intensity coefficients are

$$p_1^2 = |\mathbf{c}|^2 + \sqrt{|\mathbf{c}|^4 - |\mathbf{c}^T \mathbf{G} \mathbf{c}|^2}, \ \ p_2^2 = |\mathbf{c}|^2 - \sqrt{|\mathbf{c}|^4 - |\mathbf{c}^T \mathbf{G} \mathbf{c}|^2}. \quad (20)$$

The polarizance-diattenuation $D$ of the pure medium represented by $\mathbf{c}$ is

$$P = D = \sqrt{1 - \frac{|\mathbf{c}^T \mathbf{G} \mathbf{c}|^2}{|\mathbf{c}|^4}} = \sqrt{\frac{|\mathbf{c}^* \bullet \mathbf{c}|^2}{|\mathbf{c}|^4} - 1}, \quad (21)$$

while the degree of spherical purity is given by

$$P_S = \frac{1}{\sqrt{3}} \sqrt{1 + \frac{|\mathbf{c}^T \mathbf{G} \mathbf{c}|^2}{|\mathbf{c}|^4}} = \sqrt{3 - \frac{2|\mathbf{c}^* \bullet \mathbf{c}|^2}{|\mathbf{c}|^4}}. \quad (22)$$

The passivity condition has the expression

$$|\mathbf{c}|^2 + \sqrt{|\mathbf{c}|^4 - |\mathbf{c}^T \mathbf{G} \mathbf{c}|^2} \leq 1. \quad (23)$$

The passive representative $\tilde{\mathbf{c}}$ of a given coherency vector $\mathbf{c}$ has the form

$$\tilde{\mathbf{c}} = \mathbf{c} \Big/ \sqrt{|\mathbf{c}|^2 + \sqrt{|\mathbf{c}|^4 - |\mathbf{c}^T \mathbf{G} \mathbf{c}|^2}}. \quad (24)$$

Retarders are characterized by

$$\mathbf{T}_R^\dagger = \mathbf{T}_R^{-1}, \ \mathbf{M}_R^T = \mathbf{M}_R^{-1}, \ \mathbf{c}_R = e^{i\phi} \operatorname{diag}(i,1,1,1) \mathbf{a}_R, \ \begin{cases} \mathbf{a}_R \in \mathbb{R}^4 \\ |\mathbf{a}_R| = 1 \end{cases} \quad (25)$$

or equivalently, $\mathbf{c}_R = e^{i\phi}(\mathbf{a} + \mathbf{G}\mathbf{a}^*)/2$ with $\mathbf{a} \in \mathbb{C}^4$ and $|\mathbf{c}_R| = 1$; therefore the generic coherency vector $\mathbf{c}_R$ of a retarder can be expressed as

$$\mathbf{c}_R = e^{i\phi} \begin{pmatrix} i \cos \Delta/2 \\ \cos 2\alpha \, \sin \Delta/2 \\ \sin 2\alpha \, \sin \Delta/2 \cos \delta \\ \sin 2\alpha \, \sin \Delta/2 \sin \delta \end{pmatrix}, \quad (26)$$

where the angles $(\alpha,\delta)$ determine the eigenstates of the elliptical retarder and $\Delta$ is the retardance [1].

Normal diattenuators are characterized by

$$\mathbf{T}_D^\dagger = \mathbf{T}_D, \quad \mathbf{M}_D^T = \mathbf{M}_D, \quad \mathbf{c}_D = e^{i\phi}\mathbf{a}_D \text{ with } \mathbf{a}_D \in \mathbb{R}^4, \quad (27)$$

or equivalently, $\mathbf{c}_D = e^{i\phi}(\mathbf{a}+\mathbf{a}^*)/2$ with $\mathbf{a} \in \mathbb{C}^4$; therefore, the generic coherency vector $\mathbf{c}_D$ of a normal diattenuator can be expressed as

$$\mathbf{c}_D = e^{i\phi}\frac{1}{2}\begin{pmatrix} p_1+p_2 \\ (p_1-p_2)\cos 2\alpha \\ (p_1-p_2)\sin 2\alpha \cos\delta \\ (p_1-p_2)\sin 2\alpha \sin\delta \end{pmatrix}, \quad \begin{cases} p_1,p_2 \in \mathbb{R} \\ 0 \leq p_2 < p_1 \leq 1 \end{cases} \quad (28)$$

where the angles $(\alpha,\delta)$ determine the eigenstates of the elliptical diattenuator and $(p_1,p_2)$ are the principal amplitude coefficients [1].

Normal pure media are characterized by the following properties of their Jones matrices, Mueller matrices and coherency vectors

$$\mathbf{T}_N^\dagger \mathbf{T}_N = \mathbf{T}_N \mathbf{T}_N^\dagger, \quad \mathbf{M}_N^T \mathbf{M}_N = \mathbf{M}_N \mathbf{M}_N^T, \\ \mathbf{c}_N^T = (c_0, c_1, pc_1, qc_1)^T \quad [p,q \in \mathbb{R}]. \quad (29)$$

A pure medium is called degenerate when it has a single eigenstate (with doubly degenerate eigenvalue), and its Jones matrix can be expressed as [1,25]

$$\mathbf{T}_g = e^{i\phi}\mathbf{T}_R \mathbf{T}_{g\gamma} \mathbf{T}_R^\dagger, \\ \mathbf{T}_{g\gamma} = \begin{pmatrix} q & \gamma e^{-i\alpha} \\ 0 & q \end{pmatrix} \quad \begin{cases} q,\gamma,\alpha \in \mathbb{R} \\ q,\gamma>0, -\pi < \alpha \leq \pi \end{cases} \quad (30)$$

where $\mathbf{T}_R$ is the Jones matrix of a retarder (unitary matrix) and $\mathbf{T}_{g\gamma}$ is the Jones matrix that corresponds to a *canonical degenerate medium*, so that the coherency vector associated with $\mathbf{T}_{g\gamma}$ has the general form [1]

$$\mathbf{c}_g = (q, 0, \gamma e^{-i\alpha}, i\gamma e^{-i\alpha}) \quad \begin{cases} q,\gamma,\alpha \in \mathbb{R} \\ q,\gamma>0, -\pi < \alpha \leq \pi \end{cases} \quad (31)$$

The polar decomposition of a pure medium as a serial combination of a retarder and a normal diattenuator, can be expressed as $\mathbf{c} = \mathbf{c}_R \bullet \mathbf{c}_D = \mathbf{c}_P \bullet \mathbf{c}_R$, where the coherency vector $\mathbf{c}_R$ of the retarder coincides in both decompositions, while the coherency vector of the diattenuator takes the two respective forms $\mathbf{c}_D$ and $\mathbf{c}_P$.

In general, the formulation of any serial decomposition of a pure medium in terms of Jones or pure Mueller matrices, is directly translatable to the coherency vector formalism by substituting the matrices by the corresponding coherency vectors and inserting the coherency product in between them.

## 7. Additive composition of coherency matrices

As with the Jones matrices, the coherency vector does not hold enough parameters to represent the sixteen independent polarimetric quantities involved in a general depolarizing system. Nevertheless, through appropriate ensemble averages of pure coherency matrices, general coherency matrices (4×4 positive semidefinite Hermitian matrices that depend on up to sixteen independent real parameters) representing depolarizing systems can be generated.

When an electromagnetic wave, with an associated Stokes vector $\mathbf{s}$ representing its fixed state of polarization, interacts linearly with a sample that exhibits inhomogeneous polarimetric structure along the area transversally intersected by the wave, the sample can be considered as composed of a number $n$ of parallel components "$i$" with respective well defined coherency vectors $\mathbf{c}_i$ [i.e., with well defined Jones matrices $\mathbf{T}_i(\mathbf{c}_i)$] and respective cross-sections $p_i = A_i/A$ ($A$, $A_i$, being the total transverse area covered by the incident wave and the area of the $i$ element respectively) [15,21,26,27]. Since we are interested on considering the possible statistical fluctuations of the relative phases among the pure components, it is necessary to establish a reference for the global phase shift introduced by each element, as for instance by writing $\mathbf{c}_i = e^{i\phi_i}\hat{\mathbf{c}}_i$, with $\hat{\mathbf{c}}_i$ satisfying $\hat{\mathbf{c}}_i^\dagger \mathbf{G} \hat{\mathbf{c}}_i \in \mathbb{R}$ (note that this criterion is equivalent to the choice $\det \mathbf{T} \in \mathbb{R}$ for the associated Jones matrix). In case the field reach each patch or tile $i$ affected by different respective phases (ass occurs, for instance, in speckle patterns), they can be included in $\phi_i$. Thus, for each realization (temporal or spectral) determined by a specific set of fixed coherency vectors $\hat{\mathbf{c}}_i$, the coherency matrix $\mathbf{C}_J$ of the composed medium is pure and is generated by the coherency vector $\mathbf{c} \equiv \sum_{i=1}^n \sqrt{p_i}\mathbf{c}_i$ as follows

$$\mathbf{C}_J = \left(\sum_{i=1}^n \sqrt{p_i}\mathbf{c}_i\right) \otimes \left(\sum_{i=1}^n \sqrt{p_i}\mathbf{c}_i\right)^\dagger \\ = \left(\sum_{i=1}^n \sqrt{p_i}\mathbf{c}_i\right) \otimes \left(\sum_{i=1}^n \sqrt{p_i}\mathbf{c}_i^\dagger\right) = \mathbf{X}+\mathbf{Y}, \\ \begin{cases} \mathbf{X} \equiv \sum_{i=1}^n p_i \mathbf{C}_{Ji}, \quad [\mathbf{C}_{Ji} = \hat{\mathbf{c}}_i \otimes \hat{\mathbf{c}}_i^\dagger] \\ \mathbf{Y} \equiv \sum_{\substack{i,j=1 \\ i \neq j}}^n \sqrt{p_i p_j}\left(e^{i\phi_{ij}}\mathbf{A}_{ij}+e^{-i\phi_{ij}}\mathbf{A}_{ij}^\dagger\right) \\ [\phi_{ij} \equiv \phi_i - \phi_j, \quad \mathbf{A}_{ij} \equiv \hat{\mathbf{c}}_i \otimes \hat{\mathbf{c}}_j^\dagger] \end{cases} \quad (32)$$

where $\mathbf{X}$ is the convex sum of the coherency matrices $\mathbf{C}_{Ji}$ of the components (so that, taken isolated, $\mathbf{X}$ is a nonpure coherency matrix with $1 < \text{rank}\,\mathbf{X} \leq 4$ [the effective value of $\text{rank}\,\mathbf{X}$ depending on the dimension of the subspace of $\mathbb{C}^4$ generated by the set of vectors $\mathbf{c}_i$ $(i=1,...,n)$], while the crossed term $\mathbf{Y}$ is a Hermitian matrix that is not positive semidefinite (i.e., it is indefinite, because otherwise $\text{rank}\,\mathbf{C}_J > 1$, which violates the starting hypothesis that $\mathbf{C}_J$ is a pure coherency matrix, which in turn implies $\text{rank}\,\mathbf{C}_J = 1$). Despite these properties of $\mathbf{X}$ and $\mathbf{Y}$, their sum always results in the pure coherency matrix $\mathbf{C}_J = \mathbf{X}+\mathbf{Y}$.

As described in [3,22,23,28], the material sample can exhibit temporal or spectral fluctuations of the phase differences $\phi_{ij} = \phi_i - \phi_j$ and the effective coherency matrix $\mathbf{C}$ results from an ensemble average $\langle \mathbf{C}_J \rangle$ of the form

$$\mathbf{C} = \langle \mathbf{C}_J \rangle = \langle \mathbf{X} \rangle + \langle \mathbf{Y} \rangle, \\ \langle \mathbf{X} \rangle = \sum_{i=1}^n \langle p_i \mathbf{C}_{Ji} \rangle = \sum_{i=1}^n p_i \mathbf{C}_{Ji} = \mathbf{X}, \\ \langle \mathbf{Y} \rangle = \sum_{\substack{i,j=1 \\ i \neq j}}^n \sqrt{p_i p_j}\left[\langle e^{i\phi_{ij}}\rangle \mathbf{A}_{ij} + \langle e^{-i\phi_{ij}}\rangle \mathbf{A}_{ij}^\dagger\right], \quad \mathbf{A}_{ij} \equiv \hat{\mathbf{c}}_i \otimes \hat{\mathbf{c}}_j^\dagger. \quad (33)$$

The inspection of the above expansions shows that

1) $\mathbf{C}$, obtained through an ensemble average of pure coherency matrices, is necessarily a coherency matrix with $1 \leq \text{rank}\,\mathbf{C} \leq 4$, the value of $r \equiv \text{rank}\,\mathbf{C}$ depending on the nature of the fluctuations of the phase differences $\phi_{ij}$;

2) $\mathbf{C}_{Ji} = \mathbf{c}_i \otimes \mathbf{c}_i^\dagger$ are the pure coherency matrices of the components;

3) $\mathbf{X}$, taken isolated, represents the convex sum of the coherency matrices of the components, and therefore represents the incoherent composition of them;

4) the Hermitian matrix $\mathbf{Y} = \mathbf{C} - \mathbf{X}$ is not positive semidefinite (in fact it is necessarily indefinite), so that, except for the trivial limiting case where $\mathbf{Y} = \mathbf{0}$ (i.e., all $\phi_{ij}$ are uniformly distributed or all coherency vectors $\mathbf{c}_i$ are proportional to each other), $\mathbf{Y}$ is not a coherency matrix;

5) when all $\phi_{ij}$ do not fluctuate, the sum $\mathbf{X} + \mathbf{Y}$ results in the pure coherency matrix $\mathbf{C}_J = \mathbf{c} \otimes \mathbf{c}^\dagger$ associated with the coherency vector $\mathbf{c} = \sum_{i=1}^n \sqrt{p_i}\, \mathbf{c}_i$ (i.e., this is the limiting case where the composition is fully coherent);

6) when all phase differences $\phi_{ij}$ are uniformly distributed, the resulting coherency matrix $\mathbf{C} = \mathbf{X}$ is given by the incoherent composition (convex sum) of the coherency matrices $\mathbf{C}_{Ji} = \mathbf{c}_i \otimes \mathbf{c}_i^\dagger$, so that $r = 2,3,4$ depending on the order of the subspace of $\mathbb{C}^4$ covered by the set of coherency vectors $\mathbf{c}_i$ $(i = 1,\ldots,n)$, and

7) in general, each $\phi_{ij}$ can fluctuate in an independent manner, resulting in intermediate cases where $\mathbf{Y}$ is nonzero and takes forms different to that of the fully coherent composition, and therefore the composition is partially coherent leading to a composed coherency matrix $\mathbf{C}$ with $r = 2,3,4$, depending on the dimension of the subspace covered by the coherency vectors $\mathbf{c}_i$ and on the nature of the probability density functions of the phase differences $\phi_{ij}$.

Observe that the transformation of Stokes parameters through the action of a medium represented by a coherency matrix $\mathbf{C}$ (pure or not) is given by

$$s'_i = \sum_{j=0}^3 \mathrm{tr}(\mathbf{G}_i^* \mathbf{G}_j \mathbf{C}) s_j. \tag{34}$$

## 8. Synthesis of a depolarizing Mueller matrix

The elements $m_{ij}$ of a given Mueller matrix $\mathbf{M}$ and the elements $c_{ij}$ of its associated coherency matrix $\mathbf{C}$ (pure or nonpure) can be expressed in the condensed dual forms

$$m_{ij} = \mathrm{tr}(\mathbf{G}_i^* \mathbf{G}_j \mathbf{C}), \quad c_{ij} = \frac{1}{4} \mathrm{tr}(\mathbf{G}_i^* \mathbf{G}_j \mathbf{M}). \tag{35}$$

Due to the nature of the biunivocal the relation between $\mathbf{C}$ and $\mathbf{M}$, the Mueller matrix $\mathbf{M}$ associated with the sum of a set coherency matrices $\mathbf{C}_{Ji}$ is given by the sum of the Mueller matrices $\mathbf{M}_{Ji}$ associated with $\mathbf{C}_{Ji}$ (and vice versa), and therefore the additive properties of $\mathbf{C}$ studied in Sec. 7 are translatable to the Mueller formalism. Thus, any Mueller matrix (pure or not) can be expressed as

$$\mathbf{M} = \sum_{i=1}^n p_i \mathbf{M}_{Ji} + \mathbf{Q}(\langle \mathbf{Y} \rangle), \tag{36}$$

where the elements $q_{ij}$ of the non-Mueller matrix $\mathbf{Q}(\langle \mathbf{Y} \rangle)$ are given by $q_{ij} = \mathrm{tr}(\mathbf{G}_i^* \mathbf{G}_j \langle \mathbf{Y} \rangle)$. As with $\langle \mathbf{Y} \rangle$, $\mathbf{Q}(\langle \mathbf{Y} \rangle)$ adopts different forms depending on the probability distributions of the phase differences $\phi_{ij}$, in such a manner that, 1) $\mathbf{M}$ is a pure Mueller matrix when $\phi_{ij}$ take fixed values (i.e., do not fluctuate); 2) $\mathbf{M} = \sum_{i=1}^n p_i \mathbf{M}_{Ji}$ when all $\phi_{ij}$ are uniformly distributed (i.e., $\phi_{ij} = 1/2\pi$), and 3) $\mathbf{M}$ is a nonpure Mueller matrix that is determined by $\mathbf{C}_{Ji}$ and by the specific probability distributions of $\phi_{ij}$.

Consequently, there are infinite ways to synthesize a given nonpure Mueller matrix $\mathbf{M}$, so that the maximum number of parallel components of $\mathbf{M}$ with independent associated coherency vectors is equal to $\mathrm{rank}\,\mathbf{C}(\mathbf{M}) \equiv r$, with $1 \le r \le 4$. When $\mathbf{M}$ is pure (i.e., $\mathbf{C}$ is pure), then $r = 1$, otherwise $\mathbf{M}$ admits arbitrary parallel decompositions as convex sums of $r$ pure Mueller matrices [21], including, as a particular case, the convex sum based on the Cloude decomposition (or *spectral decomposition*) of $\mathbf{C}(\mathbf{M})$ [24].

The synthesis of a Mueller matrix from the partially coherent composition of nondepolarizing parallel components has been algebraically formulated in [28] by means of *Jones generators*, while other relevant works have dealt with the cases of partial coherence in spectroscopic polarimetry [2] and partial spatial coherence polarimetry [29].

To illustrate how the synthesis of a Mueller matrix reaches the coherent and incoherent limiting cases, let us consider a rectangular probability density function defined as $p(\phi_{ij}) = 1/2\Delta$ when $-\Delta + \phi_0 \le \phi_{ij} \le \Delta + \phi_0$ and $p(\phi_{ij}) = 0$ otherwise. Then,

$$\begin{aligned}
\langle e^{i\phi_{ij}} \rangle &= \langle e^{-i\phi_{ij}} \rangle^* = \int_{-\pi}^{\pi} p(\phi_{ij}) e^{i\phi_{ij}} d\phi_{ij} \\
&= \frac{1}{2\Delta} \int_{-\Delta+\phi_0}^{\Delta+\phi_0} e^{i\phi_{ij}} d\phi_{ij} = e^{i\phi_0} \frac{\sin\Delta}{\Delta},
\end{aligned} \tag{37}$$

so that, as expected, the coherent case is retrieved when $\Delta = 0$ (i.e., the function $\mathrm{sinc}\,\Delta \equiv (\sin\Delta)/\Delta$ takes the value $\mathrm{sinc}(0) = 1$) with a fixed value $\phi_{ij} = \phi_0$ for all phase differences, and therefore the resulting coherency and Mueller matrices are pure. The incoherent case is reached when $\phi_{ij}$ are uniformly distributed in all the zonzero interval, i.e., $p(\phi_{ij}) = 1/2\pi$ and consequently $\langle e^{i\phi_{ij}} \rangle = 0$, which in turn implies $\mathbf{Q}(\langle \mathbf{Y} \rangle) = 0$ and the resulting coherency and Mueller matrices are given by the convex sum of those of the pure components respectively. Equivalent results are obtained for the more realistic truncated normal distribution

$$p(\phi_{ij}) = \frac{e^{-(\phi_{ij} - \phi_0)^2 / (2\sigma^2)}}{\sqrt{2}\,\sigma \int_{-(\pi+\phi_0)/(\sqrt{2}\sigma)}^{(\pi-\phi_0)/(\sqrt{2}\sigma)} e^{-t^2} dt}, \tag{38}$$

which becomes a Dirac delta when $\sigma \approx 0$ (fixed $\phi_{ij}$, i.e., ), while $p(\phi_{ij}) \to 1/2\pi$ when $\sigma \gg \pi$ (fully random $\phi_{ij}$ in the interval $-\pi \le \phi_{ij} \le \pi$). Therefore, as expected, the smaller the standard deviation $\sigma$, the closer to a pure Mueller matrix $\mathbf{M}(\mathbf{C}_J)$ is, while the larger $\sigma$, the smaller the contribution of $\mathbf{Q}(\langle \mathbf{Y} \rangle)$ is, in which case the system tends to be an incoherent parallel composition represented by $\mathbf{M} = \sum_{i=1}^n p_i \mathbf{M}_{Ji}$, where $\mathbf{M}_{Ji}(\mathbf{c}_i)$ are the pure Mueller matrices of the components with associated coherency vectors $\mathbf{c}_i$.

In general, each phase difference can have a particular probability density function $p(\phi_{ij})$, so that the overall parallel composition may involve a sort of mixture of coherent, incoherent and partially coherent combinations, resulting in a Mueller matrix $\mathbf{M}$ that, except for the pure case (totally coherent composition), and by virtue of the *arbitrary decomposition* [21], is in turn equivalent to infinite different possible parallel combinations of pure components.

The obtainment of a general Mueller matrix $\mathbf{M}$ as a parallel composition of pure and passive components $\mathbf{M}_{Ji}$ ensures the covariance and passivity conditions for $\mathbf{M}$ to be a well-defined Mueller matrix (*ensemble criterion* [1]). The *covariance conditions* are given by the nonnegativity of the eigenvalues of $\mathbf{C}(\mathbf{M})$ [24], while the passivity conditions are characterized as follows [27,30] in terms of either $\mathbf{C}$ or $\mathbf{M}$,

$$\begin{aligned}
\mathrm{tr}\,\mathbf{C} + \sqrt{\sum_{j=1}^3 \mathrm{tr}^2(\mathbf{G}_j \mathbf{C})} &= m_{00}(1+D) \le 1, \\
\mathrm{tr}\,\mathbf{C} + \sqrt{\sum_{i=1}^3 \mathrm{tr}^2(\mathbf{G}_i^* \mathbf{C})} &= m_{00}(1+P) \le 1,
\end{aligned} \tag{39}$$

## 9. Covariance vector formalism

Given a Jones matrix $\mathbf{T}$, its associated covariance vector $\mathbf{h}$ is defined as $\mathbf{h}^T = (t_0, t_1, t_2, t_3)/\sqrt{2}$, which in turn determines the pure covariance matrix $\mathbf{H}_J = \mathbf{h} \otimes \mathbf{h}^\dagger$. $\mathbf{h}$ and $\mathbf{H}_J$ are linked respectively to the corresponding coherency vector $\mathbf{c}$ and coherency matrix $\mathbf{C}_J = \mathbf{c} \otimes \mathbf{c}^\dagger$ by means $\mathbf{h} = \mathcal{L}^\dagger \mathbf{c}$ and $\mathbf{H}_J = \mathcal{L}^\dagger \mathbf{C}_J \mathcal{L}$, where $\mathcal{L}$ is the unitary transfer matrix introduced in Eq. (5). In analogy to the coherency product, the *covariance product* of two covariance vectors $\mathbf{u}$ and $\mathbf{v}$ is defined as

$$\mathbf{u} \circ \mathbf{v} = \begin{pmatrix} \mathbf{u}^T \mathcal{L}^T (\mathbf{G}_0 + \mathbf{G}_1) \mathcal{L} \mathbf{v} \\ \mathbf{u}^T \mathcal{L}^T (\mathbf{G}_2 - i\mathbf{G}_3) \mathcal{L} \mathbf{v} \\ \mathbf{u}^T \mathcal{L}^T (\mathbf{G}_2 + i\mathbf{G}_3) \mathcal{L} \mathbf{v} \\ \mathbf{u}^T \mathcal{L}^T (\mathbf{G}_0 - \mathbf{G}_1) \mathcal{L} \mathbf{v} \end{pmatrix}. \quad (40)$$

As with coherency vectors, given two nondepolarizing media with respective covariance vectors $\mathbf{u}$ and $\mathbf{v}$, the covariance vector of the serial combination of such media is precisely $\mathbf{w} = \mathbf{u} \circ \mathbf{v}$. Covariance and coherency products are related through $\mathbf{u} \circ \mathbf{v} = \mathcal{L}^\dagger (\mathcal{L}\mathbf{u} \bullet \mathcal{L}\mathbf{v})$, and $\mathbf{u} \bullet \mathbf{v} = \mathcal{L}(\mathcal{L}^\dagger \mathbf{u} \circ \mathcal{L}^\dagger \mathbf{v})$.

The elements $m_{ij}$ of the Mueller matrix $\mathbf{M}_J$ associated with a given covariance vector $\mathbf{h}$ can be expressed in the form

$$m_{ij} = \mathbf{h}^\dagger \left(\mathcal{L}^\dagger \mathbf{G}_i^* \mathbf{G}_j \mathcal{L}\right) \mathbf{h}, \quad (41)$$

which shows that, in analogy to what happens with the coherency vector formalism, each element $m_{ij}$ of a pure Mueller matrix is given by a specific norm of the covariance vector through the metric $\mathcal{L}^\dagger \mathbf{G}_i^* \mathbf{G}_j \mathcal{L}$. In general, for pure or depolarizing media, $m_{ij}$ and the elements $h_{ij}$ of $\mathbf{H}$ are given as follows in terms of $\mathbf{H}$ and $\mathbf{M}$ respectively

$$m_{ij} = \mathrm{tr}\left(\mathcal{L}^\dagger \mathbf{G}_i^* \mathbf{G}_j \mathcal{L} \mathbf{H}\right), \quad h_{ij} = \left[\mathrm{tr}\left(\mathbf{F}_i^* \mathbf{F}_j \mathbf{M}\right)\right]/4,$$
$$\left[\begin{array}{l} \mathbf{F}_0 = (\mathbf{G}_0 + \mathbf{G}_1)/\sqrt{2}, \quad \mathbf{F}_1 = (\mathbf{G}_2 + i\mathbf{G}_3)/\sqrt{2}, \\ \mathbf{F}_2 = (\mathbf{G}_2 - i\mathbf{G}_3)/\sqrt{2}, \quad \mathbf{F}_3 = (\mathbf{G}_0 - \mathbf{G}_1)/\sqrt{2}. \end{array}\right] \quad (42)$$

The explicit expression for $\mathbf{H}(\mathbf{M})$ can be found in [1,31].

The properties of the coherency product shown in Sec. 5 have their counterparts in the covariance vector formalism, and are summarized in Appendix I. Interesting relations between coherency vectors, covariance vectors and matrix states are summarized in Appendix II, while the expressions for each structure in terms of the other ones are listed in Appendix III (Tables I-a and I-b) where the *complex Mueller matrices* $\langle \mathbf{T} \otimes \mathbf{T}^* \rangle$ [13] are denoted by $\mathbf{W}$ (general, depolarizing) and $\mathbf{W}_J$ (nondepolarizing).

## 10. Conclusions

The mathematical representation of linear polarimetric transformations can be performed by means of different and alternative vector and matrix formulations, which obviously are interrelated each other (see Appendix III). In this work, the coherency vector representation of nondepolarizing (or pure) media is formulated in a stand-alone way, without the necessity of using auxiliary structures. To do so, a specific *coherency product* is defined that allows for calculating the coherency vector of a serial combination of pure media as the coherency product of the coherency vectors of the serial components. The transformation of Stokes vectors by means of a pure medium is represented in a simple way through the new coherency product and expressed in terms of the coherency vector of the medium and the Stokes vector of the polarization state. The properties of this formalism are explored, showing that the symmetries of the underlying polarimetric properties are reflected in a natural way in the corresponding mathematical expressions. The general case of depolarizing (or nonpure) media is addressed in terms of coherency matrices by means of the synthesis of a general coherency matrix through the coherent, partially coherent or incoherent parallel composition of pure coherency matrices defined simply as $\mathbf{C}_{Ji} = \mathbf{c}_i \otimes \mathbf{c}_i^\dagger$ in terms of the respective coherency vectors $\mathbf{c}_i$. In analogy to the coherency vector formalism, the covariance vector formalism is also developed, including a specific product for serial compositions, which allows to identify the similarities and differences between both representations. Some interesting relations of the coherency and covariance structures to the corresponding matrix state ($\mathbf{Z}$ matrix) are analyzed in Appendix II.

**Appendix I**

Summary of properties of the covariance product defined in Sec. 9.

$\forall \mathbf{u}, \mathbf{v} \in \mathbb{C}^4, \mathbf{u} \circ \mathbf{v} \in \mathbb{C}^4$,

Given $\mathbf{u}, \mathbf{v}, \mathbf{w} \in \mathbb{C}^4$, then $(\mathbf{u} \circ \mathbf{v}) \circ \mathbf{w} = \mathbf{u} \circ (\mathbf{v} \circ \mathbf{w})$,

$\forall \mathbf{h} \in \mathbb{C}^4 \; \exists \; \mathbf{e} \ni \mathbf{h} \circ \mathbf{e} = \mathbf{e} \circ \mathbf{h}$ with $\mathbf{e}^T = (1,0,0,1)/\sqrt{2}$,

$\forall \mathbf{h} \in \mathbb{C}^4 \ni \mathbf{h}^T \mathcal{L}^T \mathbf{G} \mathcal{L} \mathbf{h} \neq 0 \; \exists \; \mathbf{h}^{-1} \in \mathbb{C}^4 \ni \mathbf{h} \circ \mathbf{h}^{-1} = \mathbf{h}^{-1} \circ \mathbf{h} = \mathbf{e}$,

$\mathbf{h}^{-1} = \mathcal{L}^\dagger \mathbf{G} \mathcal{L} \mathbf{h} / \left(\mathbf{h}^T \mathcal{L}^T \mathbf{G} \mathcal{L} \mathbf{h}\right), \; (\mathbf{u} \circ \mathbf{v})^{-1} = \mathbf{v}^{-1} \circ \mathbf{u}^{-1}$,

$\mathcal{L}^\dagger \mathbf{G} \mathcal{L} (\mathbf{u} \circ \mathbf{v}) = \left(\mathcal{L}^\dagger \mathbf{G} \mathcal{L} \mathbf{v}\right) \circ \left(\mathcal{L}^\dagger \mathbf{G} \mathcal{L} \mathbf{u}\right)$,

$(\mathbf{u} \circ \mathbf{v})^{-1} = \mathbf{v}^{-1} \circ \mathbf{u}^{-1}, \; (\mathbf{u} \circ \mathbf{v})^* = \mathbf{u}^* \circ \mathbf{v}^*$,

Given $\mathbf{u}, \mathbf{v} \in \mathbb{C}^4$ and $p, q \in \mathbb{C}$, then $p\mathbf{u} \circ q\mathbf{v} = pq(\mathbf{u} \circ \mathbf{v})$,

Given $\mathbf{a}, \mathbf{b}, \mathbf{c}, \mathbf{d} \in \mathbb{C}^4$, then $(\mathbf{a} + \mathbf{b}) \circ (\mathbf{c} + \mathbf{d}) = \mathbf{a} \circ \mathbf{c} + \mathbf{a} \circ \mathbf{d} + \mathbf{b} \circ \mathbf{c} + \mathbf{b} \circ \mathbf{d}$

The covariance product "$\circ$" is non-commutative, $\mathbf{u} \circ \mathbf{v} \neq \mathbf{v} \circ \mathbf{u}$.

**Appendix II**

Summary of properties of the matrix state generator ($\mathbf{Z}$ matrix). Coherency and covariance vectors are denoted by $\mathbf{c}$ and $\mathbf{h}$ respectively, which are linked each other by the relations $\mathbf{c} = \mathcal{L} \mathbf{h}$ and $\mathbf{h} = \mathcal{L}^\dagger \mathbf{c}$. The dependences $\mathbf{Z}(\mathbf{c})$ and $\mathbf{Z}(\mathbf{h})$ should be interpreted by applying the expressions for $\mathbf{Z}$ in terms of $\mathbf{c}$ and $\mathbf{h}$ respectively.

$\mathbf{Z}(\mathbf{c}_1 \bullet \mathbf{c}_2) = \mathbf{Z}(\mathbf{c}_1) \mathbf{Z}(\mathbf{c}_2)$,

$\mathbf{Z}(\mathbf{h}_1 \bullet \mathbf{h}_2) = \mathbf{Z}(\mathbf{h}_1) \mathbf{Z}(\mathbf{h}_2)$,

$\mathbf{Z}(\mathbf{c}) = \sum_{i=0}^{3} c_i \mathbf{G}_i^*$,

$\mathbf{Z}(\mathbf{h}) = \frac{1}{\sqrt{2}} \left[ h_0 (\mathbf{G}_0^* + \mathbf{G}_1^*) + h_1 (\mathbf{G}_2^* + i\mathbf{G}_3^*) + h_2 (\mathbf{G}_2^* - i\mathbf{G}_3^*) + h_3 (\mathbf{G}_0^* - \mathbf{G}_1^*) \right]$

$\mathbf{Z}(\mathbf{c}_1) \mathbf{c}_2 = \mathbf{Z}^*(\mathbf{c}_2^*) \mathbf{c}_1 = \mathbf{c}_1 \bullet \mathbf{c}_2$, $\mathbf{M}(\mathbf{c}) = \mathbf{Z}^*(\mathbf{c}) \mathbf{Z}(\mathbf{c}) = \mathbf{Z}(\mathbf{c}) \mathbf{Z}^*(\mathbf{c})$,

$\mathbf{M}(\mathbf{h}) = \mathbf{Z}^*(\mathbf{h}) \mathbf{Z}(\mathbf{h}) = \mathbf{Z}(\mathbf{h}) \mathbf{Z}^*(\mathbf{h})$,

$\mathcal{L}^\dagger \mathbf{Z}(\mathbf{h}_1) \mathcal{L} \mathbf{h}_2 = \mathbf{h}_1 \circ \mathbf{h}_2$,

$\mathcal{L}^T \mathbf{Z}^*(\mathbf{h}_1) \mathcal{L} \mathbf{h}_2 = \mathbf{h}_1^* \circ \mathbf{h}_2$,

$\mathbf{Z}(\mathbf{c}) \mathbf{Z}(\mathbf{G}\mathbf{c}) = \mathbf{c}^T \mathbf{G} \mathbf{c}$,

$\forall \, p, q \in \mathbb{C}, \mathbf{c}_1, \mathbf{c}_2 \in \mathbb{C}^4, \; \mathbf{Z}(p\mathbf{c}_1 + q\mathbf{c}_2) = p\mathbf{Z}(\mathbf{c}_1) + q\mathbf{Z}(\mathbf{c}_2)$,

$\mathbf{Z}(\mathbf{Z}(\mathbf{c}_1) \mathbf{c}_2) \mathbf{c}_1 = \mathbf{Z}(\mathbf{c}_1) \mathbf{Z}(\mathbf{c}_2)$,

$\mathbf{Z}(\mathbf{Z}^*(\mathbf{c}_2^*) \mathbf{c}_1) = \mathbf{Z}(\mathbf{c}_1) \mathbf{Z}(\mathbf{c}_2)$,

$\mathbf{Z}(\mathbf{G}_i^* \mathbf{c}) = \mathbf{G}_i^* \mathbf{Z}(\mathbf{c}) \; (i = 0,1,2,3)$,

$\mathbf{Z}(\mathbf{G}_i \mathbf{c}) = \mathbf{Z}(\mathbf{c}) \mathbf{G}_i^* \; (i = 0,1,2,3)$,

$\mathbf{Z}(\mathbf{c}_1) \mathbf{Z}^*(\mathbf{c}_2) = \mathbf{Z}^*(\mathbf{c}_2) \mathbf{Z}(\mathbf{c}_1)$.

**Appendix III**

Summary of the explicit relations among the different vector and matrix structures representing the polarimetric properties of material media. Notation: $\mathbf{T}$, Jones matrix; $\mathbf{h}$, covariance vector; $\mathbf{H}$, covariance matrix; $\mathbf{c}$, coherency vector; $\mathbf{C}$, coherency matrix; $\mathbf{W}$, complex Mueller matrix; $\mathbf{Z}$ matrix state, and $\mathbf{M}$, Mueller matrix. Wherever appropriate, the subscript $J$ is used for matrices to indicate that the corresponding relation is only valid for pure (nondepolarizing) matrices. Each row contains the expressions of a given structure in terms of the other ones. Notations for the elements of each structure are indicated in the boxes where the row and column categories coincide. Expressions that are valid even for depolarizing media are indicated in blue color.

The following auxiliary matrices $\mathbf{G}_i$ and $\mathbf{F}_i$ $(i=0,1,2,3)$ are used for some expressions,

$$\mathbf{G}_0 \equiv \begin{pmatrix} 1 & 0 & 0 & 0 \\ 0 & 1 & 0 & 0 \\ 0 & 0 & 1 & 0 \\ 0 & 0 & 0 & 1 \end{pmatrix}, \mathbf{G}_1 \equiv \begin{pmatrix} 0 & 1 & 0 & 0 \\ 1 & 0 & 0 & 0 \\ 0 & 0 & 0 & i \\ 0 & 0 & -i & 0 \end{pmatrix}, \mathbf{G}_2 \equiv \begin{pmatrix} 0 & 0 & 1 & 0 \\ 0 & 0 & 0 & -i \\ 1 & 0 & 0 & 0 \\ 0 & i & 0 & 0 \end{pmatrix}, \mathbf{G}_3 \equiv \begin{pmatrix} 0 & 0 & 0 & 1 \\ 0 & 0 & i & 0 \\ 0 & -i & 0 & 0 \\ 1 & 0 & 0 & 0 \end{pmatrix},$$

$$\mathbf{F}_0 = (\mathbf{G}_0 + \mathbf{G}_1)/\sqrt{2}, \quad \mathbf{F}_1 = (\mathbf{G}_2 + i\mathbf{G}_3)/\sqrt{2}, \quad \mathbf{F}_2 = (\mathbf{G}_2 - i\mathbf{G}_3)/\sqrt{2}, \quad \mathbf{F}_3 = (\mathbf{G}_0 - \mathbf{G}_1)/\sqrt{2}.$$

| | **T** | **h** | **H** | **c** |
|---|---|---|---|---|
| **T** | $\mathbf{T} \equiv \begin{pmatrix} t_0 & t_1 \\ t_2 & t_3 \end{pmatrix}, \mathbf{t} \equiv \begin{pmatrix} t_0 \\ t_1 \\ t_2 \\ t_3 \end{pmatrix}$ | $\mathbf{t} = \sqrt{2}\mathbf{h}$ | $\mathbf{t} = e^{i\phi}\sqrt{2h_{00}}\begin{pmatrix} 1 \\ h_{11}/h_{01} \\ h_{22}/h_{02} \\ h_{33}/h_{03} \end{pmatrix}$ | $\mathbf{t} = \sqrt{2}\mathcal{L}^\dagger \mathbf{c}$ |
| **h** | $\dfrac{1}{\sqrt{2}}\mathbf{t}$ | $\mathbf{h} \equiv \begin{pmatrix} h_0 \\ h_1 \\ h_2 \\ h_3 \end{pmatrix}$ | $e^{i\phi}\sqrt{h_{00}}\begin{pmatrix} 1 \\ h_{11}/h_{01} \\ h_{22}/h_{02} \\ h_{33}/h_{03} \end{pmatrix}$ | $\mathcal{L}^\dagger \mathbf{c}$ |
| **H** | $(\mathbf{t} \otimes \mathbf{t}^\dagger)/2$ | $\mathbf{h} \otimes \mathbf{h}^\dagger$ | $h_{ij}\ (i,j=0,1,2,3)$ | $\mathcal{L}^\dagger (\mathbf{c} \otimes \mathbf{c}^\dagger) \mathcal{L}$ |
| **c** | $\dfrac{1}{\sqrt{2}}\mathcal{L}\mathbf{t}$ | $\mathcal{L}\mathbf{h}$ | $e^{i\phi}\sqrt{h_{00}}\mathcal{L}\begin{pmatrix} 1 \\ h_{11}/h_{01} \\ h_{22}/h_{02} \\ h_{33}/h_{03} \end{pmatrix}$ | $\mathbf{c} \equiv \begin{pmatrix} c_0 \\ c_1 \\ c_2 \\ c_3 \end{pmatrix}$ |
| **C** | $\dfrac{1}{2}\mathcal{L}(\mathbf{t} \otimes \mathbf{t}^\dagger)\mathcal{L}^\dagger$ | $\mathcal{L}\mathbf{h} \otimes \mathbf{h}^\dagger \mathcal{L}^\dagger$ | $\color{blue}{\mathcal{L}\mathbf{H}\mathcal{L}^\dagger}$ | $\mathbf{c} \otimes \mathbf{c}^\dagger$ |
| **W** | $\mathbf{T} \otimes \mathbf{T}^*$ | $2\begin{pmatrix} h_0 & h_1 \\ h_2 & h_3 \end{pmatrix} \otimes \begin{pmatrix} h_0 & h_1 \\ h_2 & h_3 \end{pmatrix}^\dagger$ | $\color{blue}{\mathcal{L}^\dagger \mathbf{M} \mathcal{L}}$ $\color{blue}{\left[m_{ij} = \operatorname{tr}(\mathcal{L}^\dagger \mathbf{G}_i^* \mathbf{G}_j \mathcal{L} \mathbf{H})\right]}$ | $2\begin{pmatrix} h_0 & h_1 \\ h_2 & h_3 \end{pmatrix} \otimes \begin{pmatrix} h_0 & h_1 \\ h_2 & h_3 \end{pmatrix}^\dagger$ $[\mathbf{h} = \mathcal{L}^\dagger \mathbf{c}]$ |
| **Z** | $\mathcal{L}(\mathbf{T} \otimes \mathbf{I}_2)\mathcal{L}^\dagger$ | $\mathcal{L}\left[\sqrt{2}\begin{pmatrix} h_0 & h_1 \\ h_2 & h_3 \end{pmatrix} \otimes \mathbf{I}_2\right]\mathcal{L}^\dagger$ | $\mathcal{L}\left[\sqrt{2h_{00}}\begin{pmatrix} 1 & \dfrac{h_{11}}{h_{01}} \\ \dfrac{h_{22}}{h_{02}} & \dfrac{h_{33}}{h_{03}} \end{pmatrix} \otimes \mathbf{I}_2\right]\mathcal{L}^\dagger$ | $\displaystyle\sum_{i=0}^{3} c_i \mathbf{G}_i^*$ |
| **M** | $\mathcal{L}(\mathbf{T} \otimes \mathbf{T}^*)\mathcal{L}^\dagger$ | $m_{ij} = \mathbf{h}^\dagger (\mathcal{L}^\dagger \mathbf{G}_i^* \mathbf{G}_j \mathcal{L})\mathbf{h}$ | $\color{blue}{m_{ij} = \operatorname{tr}(\mathcal{L}^\dagger \mathbf{G}_i^* \mathbf{G}_j \mathcal{L} \mathbf{H})}$ | $m_{ij} = \mathbf{c}^\dagger (\mathbf{G}_i^* \mathbf{G}_j)\mathbf{c}$ |

|   | **C** | **W** | **Z** | **M** |
|---|---|---|---|---|
| **T** | $\mathbf{t} = e^{i\phi}\sqrt{2c_{00}}\,\mathcal{L}^{\dagger}\begin{pmatrix}1\\c_{11}/c_{01}\\c_{22}/c_{02}\\c_{33}/c_{03}\end{pmatrix}$ | $\mathbf{t} = e^{i\phi}\sqrt{w_{00}}\begin{pmatrix}1\\w_{03}/w_{01}\\w_{30}/w_{10}\\w_{33}/w_{11}\end{pmatrix}$ | $\mathbf{t} = \sqrt{2}\,\mathcal{L}^{\dagger}\begin{pmatrix}z_{00}\\z_{01}\\z_{02}\\z_{03}\end{pmatrix}$ | $\mathbf{t} = e^{i\phi}\sqrt{\dfrac{\mathrm{tr}\,\mathbf{M}_J}{2}}\,\mathcal{L}^{\dagger}\overline{\mathbf{c}}$ $\left[\overline{c}_i = \dfrac{\mathrm{tr}(\mathbf{G}_i^*\mathbf{G}_i\mathbf{M}_J)}{\mathrm{tr}(\mathbf{G}_i\mathbf{M}_J)}\right]$ |
| **h** | $e^{i\phi}\sqrt{c_{00}}\,\mathcal{L}^{\dagger}\begin{pmatrix}1\\c_{11}/c_{01}\\c_{22}/c_{02}\\c_{33}/c_{03}\end{pmatrix}$ | $e^{i\phi}\sqrt{\dfrac{w_{00}}{2}}\begin{pmatrix}1\\w_{03}/w_{01}\\w_{30}/w_{10}\\w_{33}/w_{11}\end{pmatrix}$ | $\mathcal{L}^{\dagger}\begin{pmatrix}z_{00}\\z_{01}\\z_{02}\\z_{03}\end{pmatrix}$ | $e^{i\phi}\dfrac{\sqrt{\mathrm{tr}\,\mathbf{M}_J}}{2}\mathcal{L}^{\dagger}\overline{\mathbf{c}}$ $\left[\overline{c}_i = \dfrac{\mathrm{tr}(\mathbf{G}_i^*\mathbf{G}_i\mathbf{M}_J)}{\mathrm{tr}(\mathbf{G}_i\mathbf{M}_J)}\right]$ |
| **H** | $\mathcal{L}^{\dagger}\mathbf{C}\mathcal{L}$ | $h_{ij} = \dfrac{\mathrm{tr}(\mathbf{F}_i^*\mathbf{F}_j\mathcal{L}\mathbf{W}\mathcal{L}^{\dagger})}{4}$ | $\mathcal{L}^{\dagger}\begin{pmatrix}z_{00}\\z_{01}\\z_{02}\\z_{03}\end{pmatrix}\otimes\begin{pmatrix}z_{00}\\z_{01}\\z_{02}\\z_{03}\end{pmatrix}^{\dagger}\mathcal{L}$ | $h_{ij} = \dfrac{\mathrm{tr}(\mathbf{F}_i^*\mathbf{F}_j\mathbf{M})}{4}$ |
| **c** | $e^{i\phi}\sqrt{c_{00}}\begin{pmatrix}1\\c_{11}/c_{01}\\c_{22}/c_{02}\\c_{33}/c_{03}\end{pmatrix}$ | $e^{i\phi}\sqrt{\dfrac{w_{00}}{2}}\mathcal{L}\begin{pmatrix}1\\w_{03}/w_{01}\\w_{30}/w_{10}\\w_{33}/w_{11}\end{pmatrix}$ | $\begin{pmatrix}z_{00}\\z_{01}\\z_{02}\\z_{03}\end{pmatrix}$ | $c_i = e^{i\phi}\dfrac{\sqrt{\mathrm{tr}\,\mathbf{M}_J}}{2}\dfrac{\mathrm{tr}(\mathbf{G}_i^*\mathbf{G}_i\mathbf{M}_J)}{\mathrm{tr}(\mathbf{G}_i\mathbf{M}_J)}$ |
| **C** | $c_{ij}\ (i,j=0,1,2,3)$ | $c_{ij} = \dfrac{\mathrm{tr}(\mathbf{G}_i^*\mathbf{G}_j\mathcal{L}\mathbf{W}\mathcal{L}^{\dagger})}{4}$ | $\begin{pmatrix}z_{00}\\z_{01}\\z_{02}\\z_{03}\end{pmatrix}\otimes\begin{pmatrix}z_{00}\\z_{01}\\z_{02}\\z_{03}\end{pmatrix}^{\dagger}$ | $c_{ij} = \dfrac{\mathrm{tr}(\mathbf{G}_i^*\mathbf{G}_j\mathbf{M})}{4}$ |
| **W** | $\mathcal{L}^{\dagger}\mathbf{M}\mathcal{L}$ $\left[m_{ij} = \mathrm{tr}(\mathbf{G}_i^*\mathbf{G}_j\mathbf{C})\right]$ | $w_{ij}\ (i,j=0,1,2,3)$ | $\mathcal{L}^{\dagger}\mathbf{Z}\mathbf{Z}^{*}\mathcal{L}$ | $\mathcal{L}^{\dagger}\mathbf{M}\mathcal{L}$ |
| **Z** | $c_{00}\left[\mathbf{G}_0^* + \sum_{k=1}^{3}\dfrac{c_{kk}}{c_{0k}}\mathbf{G}_k^*\right]$ | $e^{i\phi}\sqrt{\dfrac{w_{00}}{2}}\sum_{i=0}^{3}\left[\mathcal{L}\begin{pmatrix}1\\w_{03}/w_{01}\\w_{30}/w_{10}\\w_{33}/w_{11}\end{pmatrix}\right]_i \mathbf{G}_i^*$ | $z_{ij}\ (i,j=0,1,2,3)$ | $e^{i\phi}\dfrac{\sqrt{\mathrm{tr}\,\mathbf{M}_J}}{2}\mathbf{Z}(\overline{\mathbf{c}})$ $\left[\overline{c}_i = \dfrac{\mathrm{tr}(\mathbf{G}_i^*\mathbf{G}_i\mathbf{M})}{\mathrm{tr}(\mathbf{G}_i\mathbf{M})}\right]$ |
| **M** | $m_{ij} = \mathrm{tr}(\mathbf{G}_i^*\mathbf{G}_j\mathbf{C})$ | $\mathcal{L}\mathbf{W}\mathcal{L}^{\dagger}$ | $\mathbf{Z}\mathbf{Z}^{*}$ | $m_{ij}\ (i,j=0,1,2,3)$ |